\documentclass[5p,sort&compress]{elsarticle}
\usepackage{textcomp}
\usepackage[utf8]{inputenc} % keeps arXiv calm about Unicode
\usepackage{amsmath,amssymb,graphicx,tabularx,lineno,textcomp,gensymb,lscape}
\usepackage{siunitx}
\DeclareSIUnit{\atomicpercent}{at.\%}
\DeclareSIUnit{\wtpercent}{wt.\%}
\DeclareSIUnit{\angstrom}{\text{\AA}}
\DeclareSIUnit{\h}{\hour}
\makeatletter
\@ifundefined{linenumbers}{}{}
\@ifundefined{modulolinenumbers}{}{}
\@ifundefined{nolinenumbers}{}{}
\makeatother

\usepackage{amssymb}

\begin{document}
\begin{frontmatter}
\title{Time-Dependent Oxidation and Scale Evolution of a\\ Wrought Co/Ni-based Superalloy}

\author[IC]{Cameron Crabb*}\ead{c.crabb23@ic.ac.uk}
\author[IC]{Zachary T. Kloenne}
\author[IC]{Samuel R. Rogers}
\author[IC]{Chi Hang D. Kwok}
\author[RR]{Mark C. Hardy}
\author[IC]{Michele S. Conroy}
\author[IC]{David Dye*}\ead{ddye@ic.ac.uk}

\address[IC]{Department of Materials, Royal School of Mines, Imperial College London, Prince Consort Road, London, SW7 2BP, UK}
\address[RR]{Rolls-Royce plc., Elton Road, Derby, DE24 8BJ, UK}

\begin{abstract}
\noindent Understanding how protective oxide scales evolve over time is necessary for improving the long term resistance of superalloys. This work investigates the time-dependent oxidation behavior of an ingot-processable Co/Ni-based superalloy oxidized in air at \SI{800}{\degreeCelsius} for \SIlist{20;100;1000}{\hour}. Mass-gain and white-light interferometry measurements quantified oxidation kinetics, surface roughness, and spallation, while high-resolution STEM-EDX characterized oxide morphology and nanoscale elemental partitioning. Atom probe tomography captured the key transition regions between the chromia and alumina scales, and X-ray diffraction was used to identify a gradual transition from NiO and (Ni,Co)-spinel phases to a compact, dual phase chromia and alumina-rich scale. The oxidation rate evolved from near-linear to parabolic behavior with time, consistent with diffusion-controlled growth once a continuous Cr$_2$O$_3$/$\alpha$-Al$_2$O$_3$ scale formed. These observations help link kinetics, structure and chemistry, showing how an originally porous spinel layer transforms into a dense, adherent chromia + alumina scale that provides long-term protection in wrought Co/Ni-based superalloys.
\end{abstract}

\end{frontmatter}
\setpagewiselinenumbers

\section{Introduction}
Superalloys are used in high-temperature components where both mechanical strength and oxidation resistance are essential for operation, namely the hottest sections of jet engines, such as the turbine \cite{Reed2006TheApplications}. At these high temperatures environmental degradation, such as oxidation and corrosion resistance, can become life-limiting. The environmental performance of an alloy depends on it's ability to form a slow-growing, adherent oxide scale that restricts further internal oxidation \cite{Bose2017HighCoatings,Klein2013TheAlloys,Giggins1971Oxidation1200C}. In Ni-based alloys, protection mainly arises from the development of continuous Al$_2$O$_3$ or Cr$_2$O$_3$ layers via selective oxidation that limits oxygen ingress. Giggins and Pettit defined three oxidation regimes in Ni--Cr--Al systems: (I) non-protective NiO formation, (II) mixed Cr$_2$O$_3$ with internal Al$_2$O$_3$, and (III) fully protective external Al$_2$O$_3$ \cite{Giggins1971Oxidation1200C}. Once a continuous Al$_2$O$_3$ or Cr$_2$O$_3$ layer forms, oxidation becomes diffusion-limited and follows parabolic kinetics, whereas earlier stages are typically linear and cation controlled \cite{Giggins1971Oxidation1200C,Wagner1952TheoreticalAlloys,Prescott1992TheAlloys,Liang2021InfluenceSuperalloys}. The transition between these regimes depends on temperature, exposure time, and composition: around \SIrange{5}{8}{\atomicpercent} Al and about \SI{10}{\atomicpercent} Cr has been shown to promote protective behavior, while Ti has been shown to accelerate NiO formation and reduce scale adhesion \cite{Ma2022TheSuperalloys,Park2015EffectsSuperalloys,Yang1981EffectSuperalloy}. Small additions of refractory elements such as Ta or Hf can stabilize the alumina, although their effects are composition-dependent \cite{Donachie2002Superalloys:Guide,Klein2011HighSuperalloys,Yang1981EffectSuperalloy,Park2015EffectsSuperalloys,Ye2021InfluenceSuperalloy,Barrett1983The1150C}.  

Cobalt-based alloys exhibit worse oxidation resistance than Ni-based because CoO and Co$_3$O$_4$ are non-protective and volatile at elevated temperatures \cite{Klein2011HighSuperalloys,Yan2012QuaternarySuperalloys,Sato2006Cobalt-baseAlloys}. The development of $\gamma'$ strengthened Co--Al--W alloys has renewed interest in the use of Co-based alloys in industry, but stable oxidation protection still required sufficient Al and Cr to form layered chromia--alumina scales \cite{Zhang2022RevealingSuperalloys,Li2022EffectsSuperalloys}. This helped lead to the addition of Ni to the Co solvent leading to dual Co/Ni systems that combine the high temperature strength of Co alloys with the oxidation performance typical of Ni alloys, stabilizing mixed Cr$_2$O$_3$/Al$_2$O$_3$ structures and suppressing Co-oxide formation \cite{Knop2014ASuperalloy, Klein2011HighSuperalloys,Yan2014AlloyingAlloys,Li2022EffectsSuperalloys,Gao2019HighAddition}. At \SIrange{800}{900}{\degreeCelsius}, these alloys commonly evolve from spinel-dominated scales to protective mixed chromia and alumina layers, though over longer timescales than conventional Ni-base counterparts \cite{Klein2011HighSuperalloys,Yan2014AlloyingAlloys}. The interactions among Co, Ni, and reactive or refractory elements remain complex: Ti and Nb can either improve or degrade oxidation behavior depending on local chemistry, while Cr is vital to stable scale formation \cite{Ye2021InfluenceSuperalloy,Yan2014AlloyingAlloys,Zhang2024EffectSuperalloys}.  

\begin{table*}[t!]
\centering
 \caption{Nominal composition in \si{\wtpercent} of alloy chosen for the experiments \cite{Hardy2022EP003957761A1GB}.}
  \begin{tabular}[htbp]{@{}lllllllllllllllll@{}}
    \hline
    \textbf{Element} & \textbf{Ni} & \textbf{Co} & \textbf{Cr} & \textbf{W} & \textbf{Mo} & \textbf{Al} & \textbf{Ta}& \textbf{Nb} & \textbf{Ti} & \textbf{C} & \textbf{B} & \textbf{Zr} \\
    \hline
     \si{\wtpercent} & 50.8 & 19.8 & 10.5  & 6.2 & 2.8 & 4.4 & 3.8 & 0.8 & 0.8 & 0.015 & 0.009 & 0.03  \\
    \hline
  \end{tabular}
\end{table*}

Understanding these regimes requires correlating oxidation kinetics with nanoscale structure and chemistry. The present study investigates oxidation in a wrought Co/Ni-based superalloy exposed in air at \SI{800}{\degreeCelsius} for \SIlist{20;100;1000}{\hour} to represent the early, transitional, and steady-state regimes described above. Mass-gain and white-light interferometry (WLI) help establish oxidation kinetics and surface evolution; high-resolution scanning transmission electron microscopy coupled with energy-dispersive X-ray spectroscopy (STEM-EDX) characterise the scale morphology and stratification; and atom probe tomography (APT) at \SI{20}{\hour} helps resolve the atomic-scale chemistry of the transient scale. X-ray diffraction (XRD) investigates the phase evolution from (Ni,Co)-spinel to a mixture of chromia and alumina. These complementary techniques link the typical oxidation regimes to both structural and chemical observations, demonstrating how a transient, porous spinel evolves into a dense, adherent Cr$_2$O$_3$/$\alpha$-Al$_2$O$_3$ composite scale that provides long-term oxidation resistance at \SI{800}{\degreeCelsius} \cite{Liu2010OxidationAir,Rodenkirchen2022ASuperalloys,Wo2024TheSuperalloy,Blavette2000TheSuperalloys}.

\section{Experimental Description}
\subsection{Oxidation Experiment}

The investigated alloy was a wrought Co/Ni-based superalloy that was developed from the earlier V208 series by Knop et al. \cite{Knop2014ASuperalloy,Hardy2022EP003957761A1GB}. Its nominal composition is given in \textbf{Table~1}. Rectangular specimens measuring \SI{15}{\milli\metre}~\(\times\)~\SI{15}{\milli\metre} were mechanically polished to an in-service surface finish following standard metallographic procedures. Oxidation tests were performed in a laboratory furnace at \SI{800}{\degreeCelsius}, with the temperature confirmed with a thermocouple. Tests were performed over three time-scales: \SIlist{20;100;1000}{\hour} to capture early, transitional, and steady-state oxidation regimes.  

Mass-gain measurements were taken using a Mettler Toledo DSC/TGA 2 instrument under static dry air to validate kinetic behavior. The mass gain per unit area and its square were plotted as functions of time to gain insight into oxidation rate laws and deviations from parabolic growth.

\subsection{Characterisation}

XRD was performed using a Bragg--Brentano diffractometer equipped with a Cu K$\alpha$ radiation source ($\lambda = \SI{1.5418}{\angstrom}$). Scans were collected over a 2$\theta$ range of \SIrange{20}{80}{\degree} with a step size of \SI{0.02}{\degree} and a counting time of \SI{1}{\second} per step. The instrument was operated at \SI{30}{\kilo\volt} and \SI{10}{\milli\ampere}, where the peaks were matched against the ICDD PDF-5+ database. Surface topography was characterized by WLI to determine the average surface roughness, \textit{S\textsubscript{a}}, extracted from the reconstructed surface profiles.

Site-specific lamellae for STEM and APT were extracted using a standard in situ lift-out procedure using a Thermo Fisher Helios CX dual-beam focused ion beam–scanning electron microscope (FIB-SEM). Final polishing was performed at \SI{5}{\kilo\volt} and \SI{2}{\kilo\volt} to minimize surface damage and gallium implantation. All STEM datasets were acquired using a probe-corrected Thermo Fisher Spectra 300 microscope operating at \SI{300}{\kilo\volt}, equipped with a Bruker Dual-X energy dispersive X-ray (EDX) detector. The convergence semi-angle was approximately \SI{30}{\milli\radian}. The probe current used was \SI{1}{\nano\ampere}. 

APT analysis was performed under conditions of \SI{50}{\kilo\volt} and \SI{50}{\pico\joule} using a CAMECA LEAP 5000XR with a pulse frequency of \SI{250}{\kilo\hertz} and a detection rate of \SI{1}{\percent}. The shank angle of each specimen was taken from SEM images prior to analysis and applied for three-dimensional reconstruction. This allowed increased confidence in the correlation between atomic-scale chemistry and microstructural features observed by STEM-EDX.  

Together, these complementary techniques—mass-gain, TGA, XRD, WLI, STEM-EDX, and APT—provided a multiscale dataset linking oxidation kinetics, surface evolution, and nanoscale chemistry across the three oxidation time intervals.

\section{Results}

\subsection{General Oxidation Behavior}
\textbf{Figure~1} shows the specific mass gain as a function of time for specimens oxidized in air at \SI{800}{\degreeCelsius} up to \SI{100}{\hour}. It can be seen that up to \SI{20}{\hour} the alloy exhibits an expected near-linear increase, while by \SI{100}{\hour} it approaches near linearity, consistent with diffusion-limited growth.

\begin{figure}[b!]
\centering
 \includegraphics[width=0.8\columnwidth]{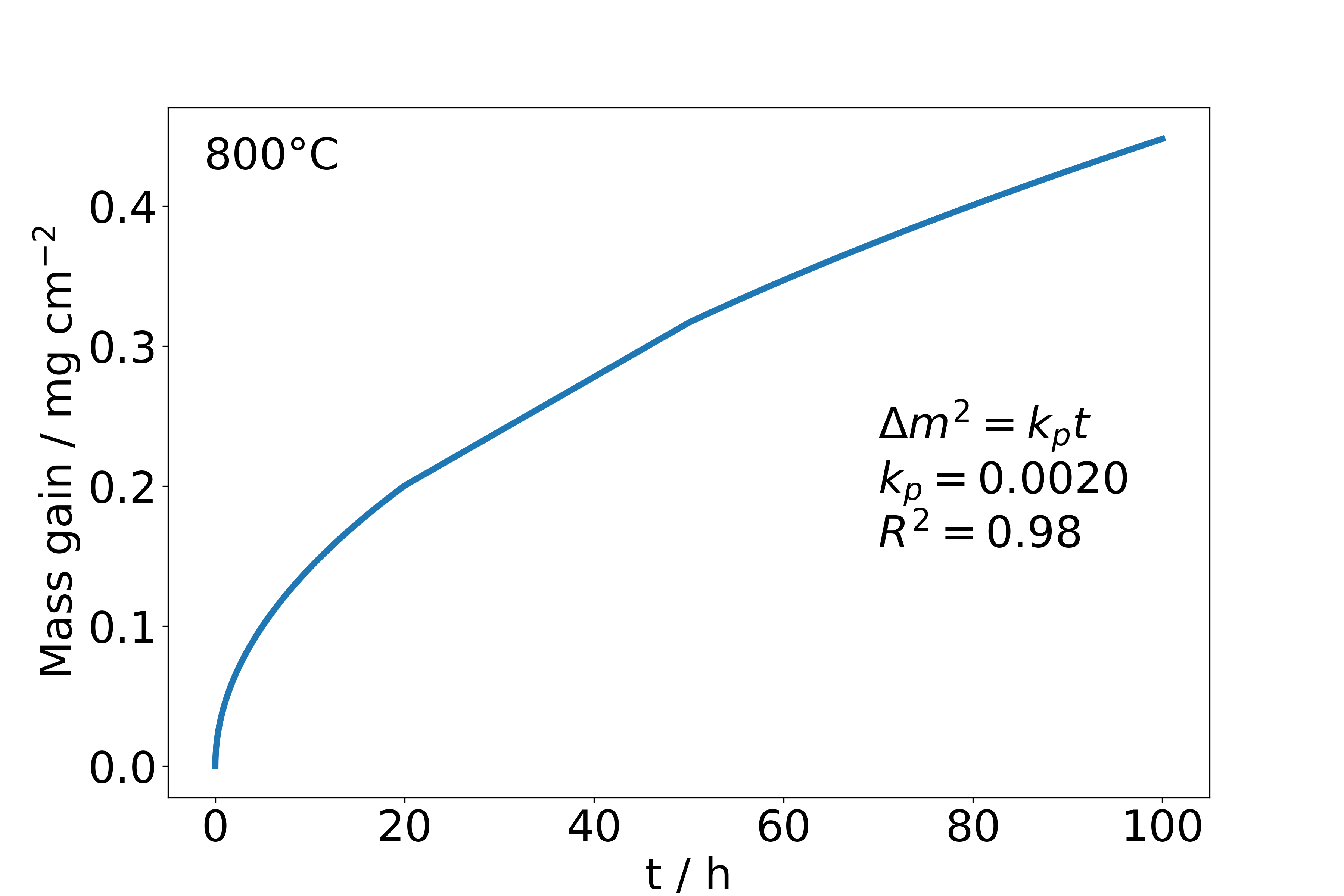}
  \caption{Mass gain versus oxidation time highlighting the isothermal oxidation behavior of the investigated alloy in static, dry air at \SI{800}{\degreeCelsius} for the first \SI{100}{\hour} of exposure.}
  \label{fig:massgaingraph}
\end{figure}

\begin{figure}[t!]
\centering
 \includegraphics[width=\columnwidth]{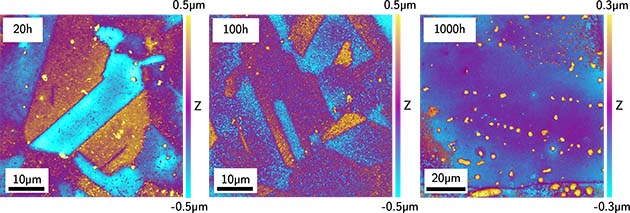}
  \caption{White light interferometry images of the chosen alloy investigated in isothermal oxidation at \SI{800}{\degreeCelsius} for \SIlist{20;100;1000}{\hour}.}
  \label{fig:WLI}
\end{figure}

\begin{figure}[t!]
\centering
 \includegraphics[width=\columnwidth]{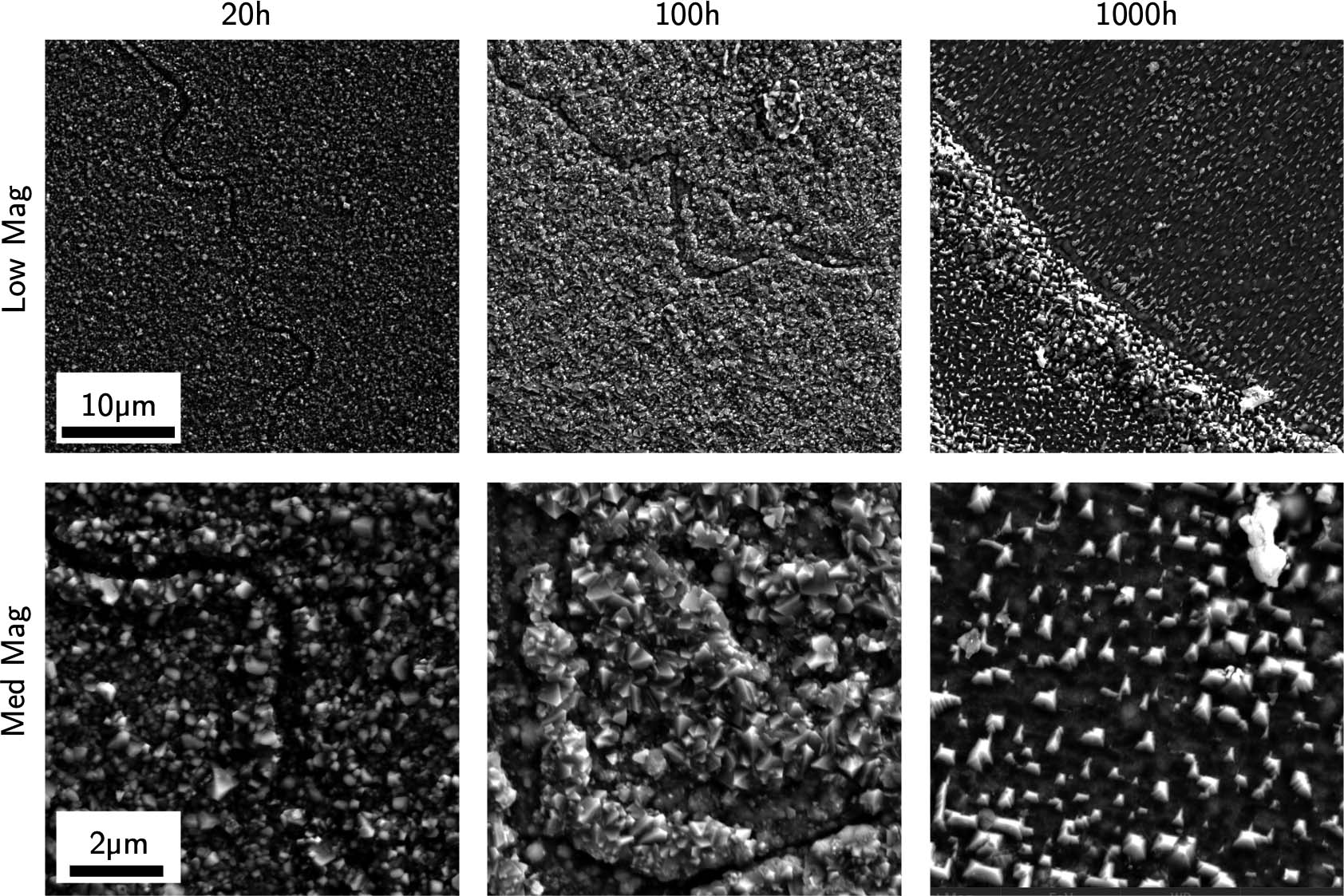}
  \caption{Secondary electron images highlighting surface roughness of the oxidised surfaces at \SI{800}{\degreeCelsius} after \SIlist{20;100;1000}{\hour}.}
  \label{fig:surfaceSEM}
\end{figure}

Surface topography measured by WLI and SEM are highlighted in \textbf{Figures~\ref{fig:WLI}--\ref{fig:surfaceSEM}}. At \SI{20}{\hour} and \SI{100}{\hour} there is a clear roughness on the surface of the grains and an apparent depression at the grain boundaries, consistent with a developing scale rich in spinel oxides; by \SI{1000}{\hour} the surface has smoothed out highlighting the removal of these spinel oxides via spallation. Grain boundaries also seem to exhibit a limited depression which suggests oxide overgrowth, enhancing protection. WLI maps help quantify these observations, shown in \textbf{Figure~3}. After \SI{20}{\hour} the average surface roughness ($S_\mathrm{a}$) was $S_\mathrm{a} = \SI{0.158}{\micro\metre}$ with a clear depression at the grain boundary. Interestingly, depression associated with twin behavior is also observed indicating a twin-related oxidation mechanism. Similar behavior is observed after \SI{100}{\hour}, with $S_\mathrm{a} = \SI{0.105}{\micro\metre}$ and the same grain-boundary depressions indicating slower local oxidation in these regions. After \SI{1000}{\hour} the overall surface is noticeably smoother with $S_\mathrm{a} = \SI{0.076}{\micro\metre}$. However, isolated bulging features are present near the grain boundaries. Across the series there is an apparent reduction in visible grain boundary depression, suggesting progressive overgrowth and potential grain-boundary grooving and/or infilling by the maturing oxide scale, protecting these regions.

\subsection{Cross-Sectional Morphology and Phase Constitution}
High angular dark field (HAADF) images of the oxide cross-sections are shown in \textbf{Figure~4}. At \SI{20}{\hour} a multilayered, rough, porous oxide scale is observed with a thickness of \SI{1}{\micro\metre}. A discontinuous layer of acicular intrusions can be seen, disconnected from the bulk oxide. By \SI{100}{\hour}, a less rough, more compact scale is evident with a thickness of \SI{0.5}{\micro\metre}, and some preferential growth is also observed down the grain boundary. At \SI{1000}{\hour}, a much more densely compact multilayer is observed with smoother morphology and reduced porosity. The compact oxide layer extends to about \SI{0.5}{\micro\metre}. The acicular layer appears more well connected; however, internal precipitates can be observed and the overall oxide damage layer depth is observed to be \SI{3.4}{\micro\metre}. 
\begin{figure}[t!]
\centering
 \includegraphics[width=\columnwidth]{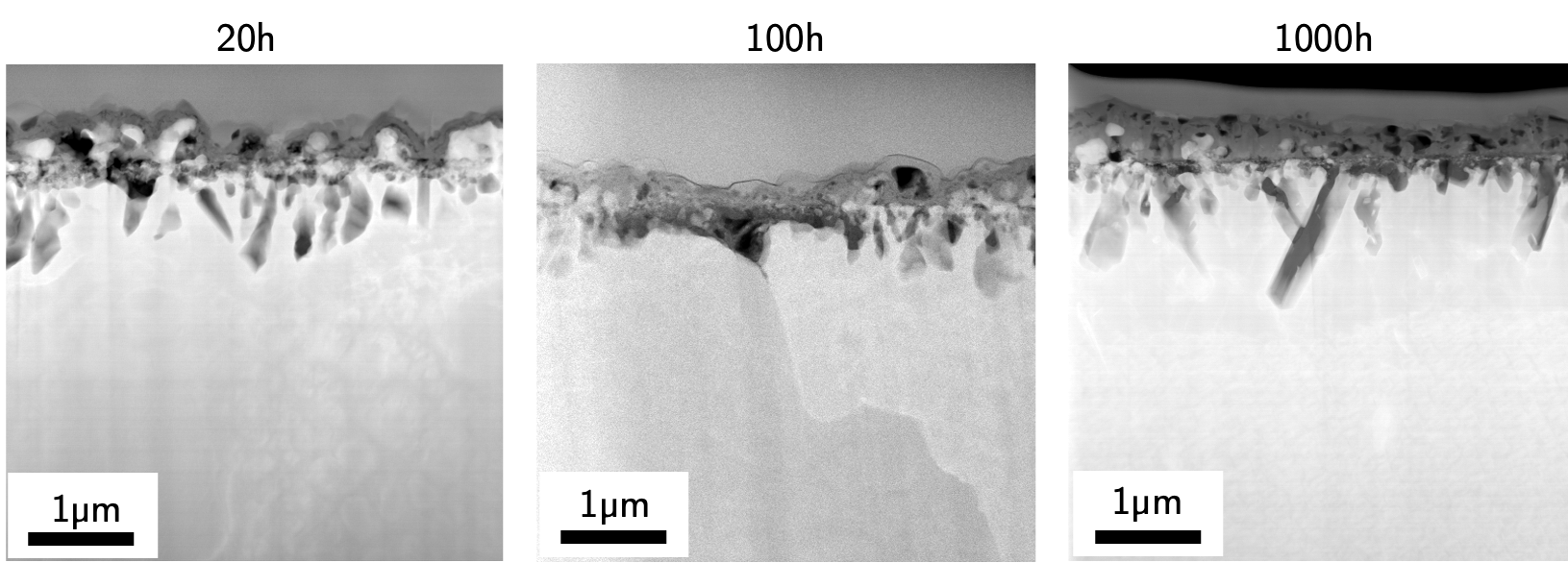}
  \caption{HAADF images taken at each time interval, highlighting scale development.}
  \label{fig:HAADF}
\end{figure}

XRD patterns for the three exposure times are shown in \textbf{Figure~5}. After \SI{20}{\hour}, the diffractogram is dominated by the substrate ($\gamma$/$\gamma$$'$) reflections, with weak contributions from spinel, Al$_2$O$_3$ and Cr$_2$O$_3$, consistent with the onset of a transient outer scale. By \SI{100}{\hour}, peaks from (Ni,Co)Cr$_2$O$_4$ spinel and Cr$_2$O$_3$ intensify, confirming the development of a duplex oxide with a continuous chromia layer. At \SI{1000}{\hour}, spinel reflections are markedly reduced, while Cr$_2$O$_3$ remains prominent and new peaks corresponding to $\alpha$-Al$_2$O$_3$ appear, indicating the establishment of a dense, protective chromia–alumina bi-layer. Weak rutile (TiO$_2$) reflections are also present at intermediate and long exposures. The apparent reduction in overall peak intensities at \SI{1000}{\hour} suggests increased X-ray absorption and/or preferred orientation within the compact $\alpha$-Al$_2$O$_3$ + Cr$_2$O$_3$ scale, potentially compounded by peak broadening from fine $\alpha$-Al$_2$O$_3$ grains. Overall, oxidation at \SI{800}{\degreeCelsius} evolves from a transient spinel-rich scale to a chromia-dominated structure capped by $\alpha$-Al$_2$O$_3$, with alumina increasingly contributing to long-term protection.

\begin{figure}[t!]
\centering
 \includegraphics[width=\columnwidth]{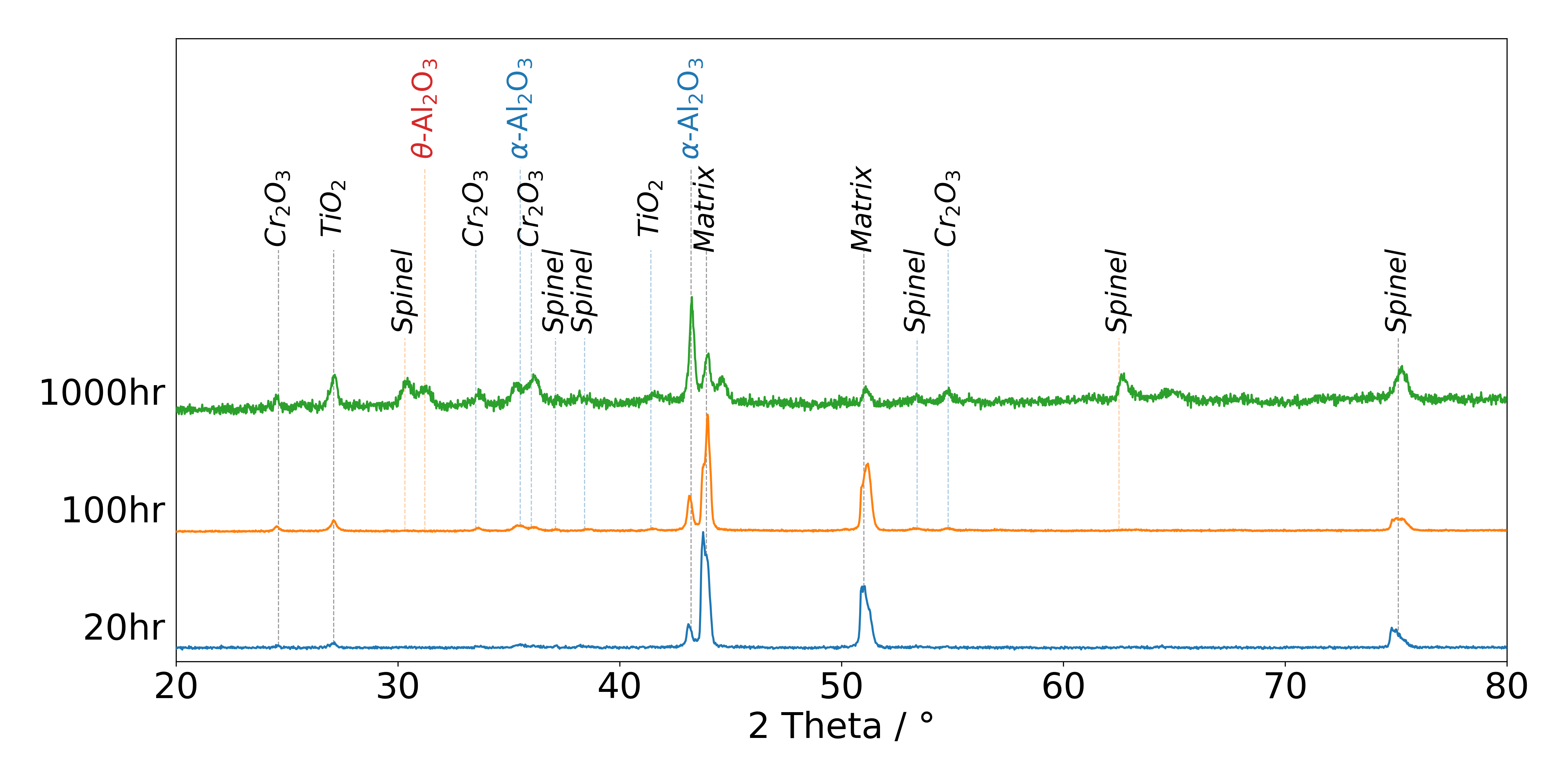}
  \caption{XRD graph highlighting the time-dependent phase development of the investigated alloy under isothermal oxidation at \SI{800}{\degreeCelsius}.}
  \label{fig:oxidationgraph}
\end{figure}

\subsection{Nanoscale Characterization}
High-resolution STEM-EDX maps of the oxide scale after \SI{20}{\hour} and \SI{1000}{\hour} exposures are shown in \textbf{Figures~6--7}. At \SI{20}{\hour}, the oxide is irregular and rough, with the outermost regions dominated by Co- and Ni-rich phases, corresponding to typical spinel oxides. Beneath this, a continuous Cr-rich layer is already established. For both Cr and Ni a compositional gradient is observed in the bulk metal, in which Cr becomes depleted toward the metal/oxide interface while Ni shows a slight enrichment in the same region. Bands enriched in Ti and Ta are evident, corresponding respectively to rutile (TiO$_2$) and Ta$_2$O$_5$, accompanied by detectable Nb, Mo, and W signals. Numerous bubble-like features enriched in Ni, Co, and refractory elements are distributed throughout the oxide scale. Below these layers, an emerging Al-enriched layer consistent with (Al$_2$O$_3$) can be observed, extending as acicular intrusions into the bulk. The underlying $\gamma'$ microstructure remains intact and closely resembles that of the heat-treated condition of the alloy prior to oxidation. 

After \SI{1000}{\hour}, the oxide has developed into a denser and more uniform structure. The layers associated with Nb, W, and Mo are no longer prominent, and most of the Ta$_2$O$_5$ layer observed at \SI{20}{\hour} has diminished. The outer spinel layer has become more continuous and compact, forming a more stable surface layer. A thick and persistent Cr$_2$O$_3$ layer dominates the scale, beneath which a uniform continuous alumina layer is present. Several of the bubble-like features observed in the \SI{20}{\hour} sample appear to have sealed within the oxide scale, still containing evidence of refractory enrichment. Ti remains locally detectable, particularly near internal precipitates within the oxide co-located with Nb in addition to the evident rutile (TiO$_2$) layer in the scale. In the underlying alloy, the $\gamma'$ distribution has evolved towards unimodal, with a clearly defined depletion region below the oxide scale in addition to the internal Ti/Nb-rich precipitates extending into the matrix.

APT was performed on site-specific needles extracted from the \SI{20}{\hour} specimen, positioned within the chromia layer and extending through one of the bubble-like features into the underlying alumina region to capture the early development of these internal oxides, \textbf{Figure~8}. The reconstruction reveals five distinct compositional regions across the oxide scale. At the surface, a CrO$^+$-rich zone corresponds to the chromia layer observed in STEM-EDX, showing a uniform chromium distribution with little evidence of other cations. Directly beneath, a narrow layer dominated by TaO$^+$ and TaO$_2^+$ ions is detected, marking a discrete Ta$_2$O$_5$ band separating the chromia from the lower oxide regions. Beneath this, a clearly defined Al- and O-enriched region is present, consistent with alumina, within which elemental Cr$^+$ ions are also detected, suggesting local incorporation of metallic Cr. Next to the alumina, a spinel and refractory-rich zone is observed that coincides with the bubble region; this area contains elevated concentrations of Ni, Co, Ta$^+$, Nb$^+$, MoO$^+$, and W, indicating the accumulation of multiple cations and complex oxide species. At the boundary between the (Al$_2$O$_3$) and the bubble, a sharp interface is noted by segregation of NbO$^+$ and WO$^+$ species, defining a clear interfacial region enriched in refractory oxides.

\begin{figure*}[p]
\centering
 \includegraphics[width=1.8\columnwidth]{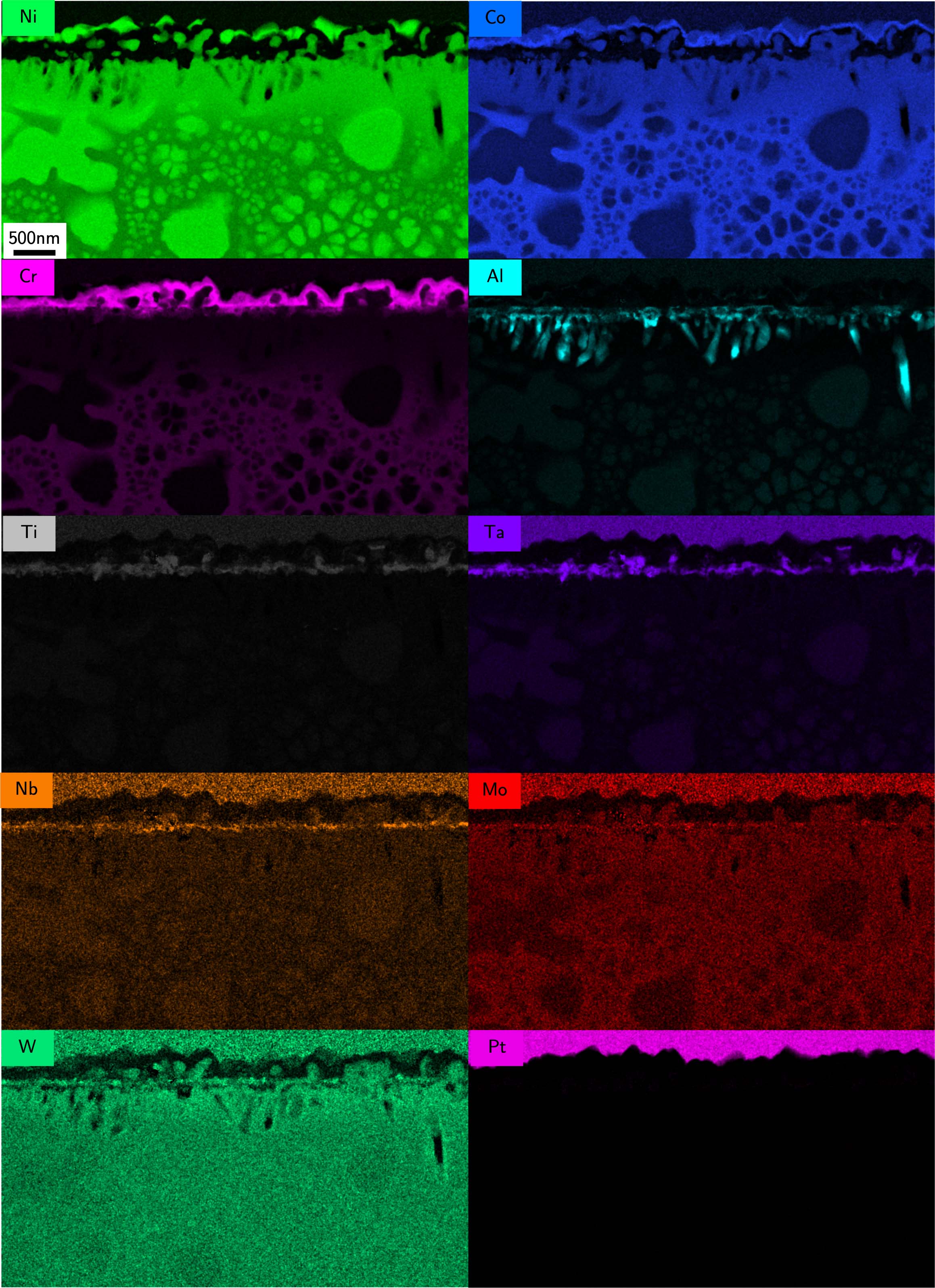}
  \caption{STEM-EDX of the cross-section of the investigated alloy after isothermal oxidation at \SI{800}{\degreeCelsius} for \SI{20}{\hour}. Maps shown in \si{\atomicpercent}.}
  \label{fig:boat1}
\end{figure*}

\begin{figure*}[p]
\centering
 \includegraphics[width=1.8\columnwidth]{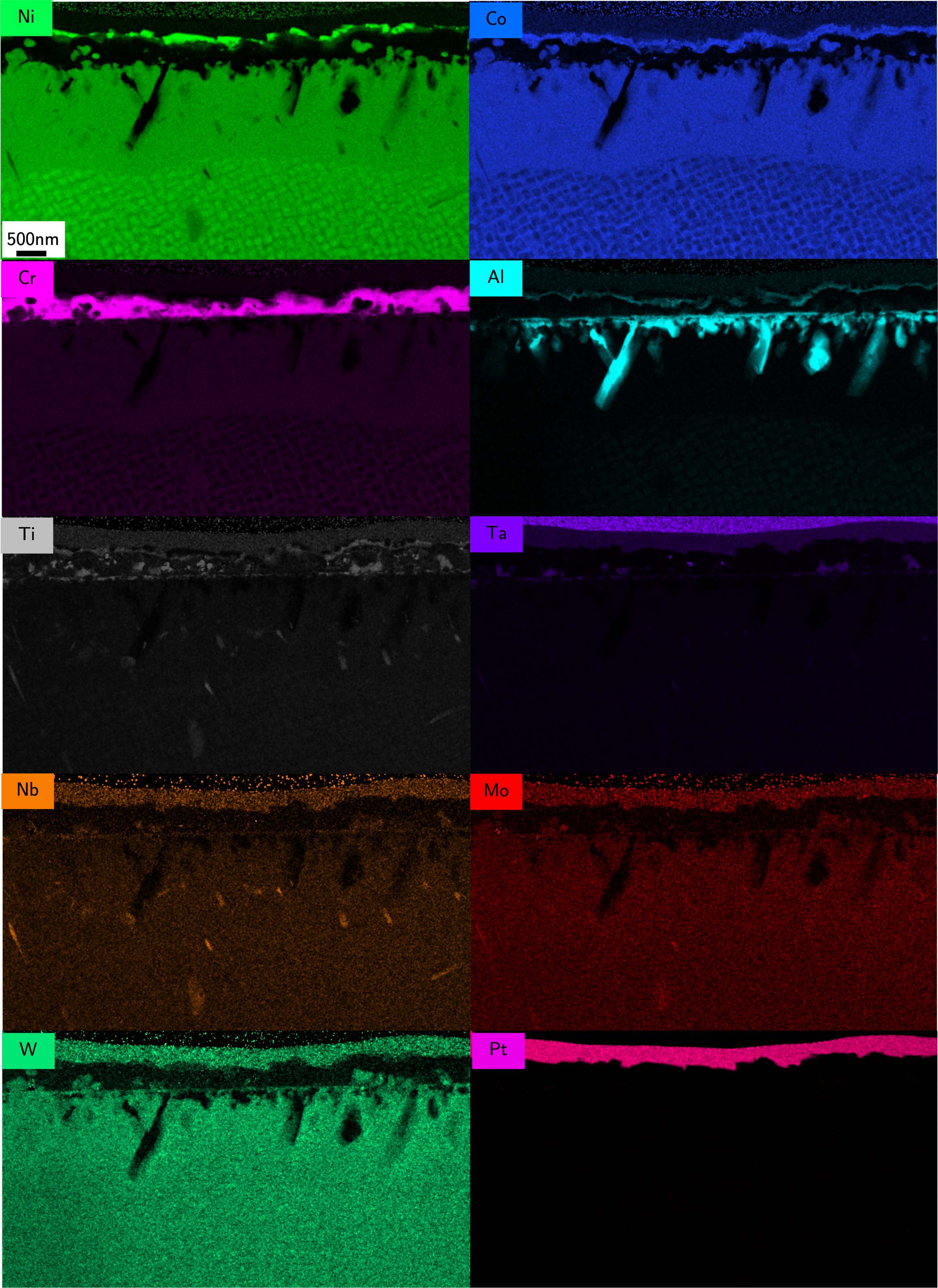}
  \caption{STEM-EDX of the cross-section of the investigated alloy after isothermal oxidation at \SI{800}{\degreeCelsius} for \SI{1000}{\hour}. Maps shown in \si{\atomicpercent}.}
  \label{fig:STEMEDX1000}
\end{figure*}

\begin{figure*}[t]
\centering
 \includegraphics[width=1.6\columnwidth]{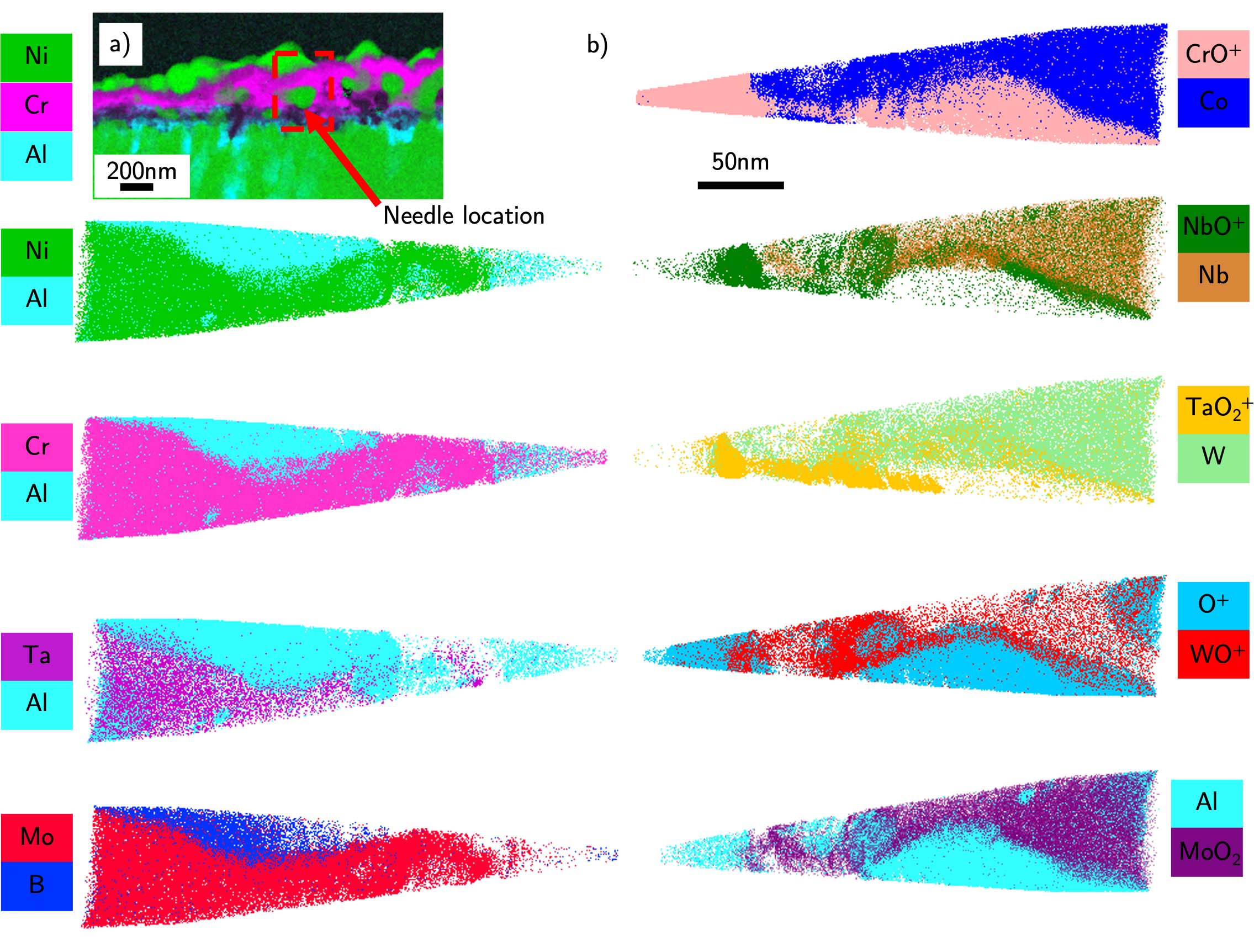}
  \caption{(a) STEM-EDX image of the APT needle lift-out location through Cr$_2$O$_3$ scale and refractory--rich spinel bubble into the Al$_2$O$_3$ fingers on \SI{20}{\hour} oxidized sample. (b) APT needle reconstructions obtained for different species.}
  \label{fig:APT}
\end{figure*}

\section{Discussion}

The oxidation behavior of this wrought Co/Ni-based superalloy at \SI{800}{\degreeCelsius} shows a clear time-dependent evolution linking surface morphology, phase constitution, and nanoscale chemistry. The transition from initial rapid oxidation to slower, diffusion-controlled growth is reflected in both the curvature of the mass-gain curves and the surface roughness reduction across the time scales. The near-linear kinetics at \SI{20}{\hour} correspond to cation-driven growth through a rough, spinel-rich surface layer, consistent with early-stage oxidation reported in Co- and Ni-base systems where outward cation diffusion dominates before a continuous chromia barrier forms \cite{Wang2023TheSuperalloy,Tian2025Co-evolutionProperties}. The subsequent reduction in rate coincides with the development and thickening of a dense Cr$_2$O$_3$ layer, which acts as an effective diffusion barrier. This progressive increase in stability shown by the WLI observations, showing a smoother overall surface at \SI{1000}{\hour} with limited bulging of Cr-rich oxide and reduced roughness relative to earlier exposures.

At short timescales, the alloy surface oxidizes rapidly to form a rough scale dominated by (Ni,Co)(Cr,Al)$_2$O$_4$ spinel and a chromia band that has already been established. Ni-rich regions are seen throughout the surface of the scale,  which displays depressions along grain boundaries, highlighting preferential oxidation down the grain boundaries which act as fast-diffusion pathways. Oxygen penetration down the boundaries promotes inward growth of oxide, producing the characteristic trough-like features captured by WLI. Beneath the spinel layer, a continuous Cr-rich layer with a thin alumina enriched zone beneath it can be seen, marking the onset of sub-scale alumina formation. Discrete Ti- and Ta-containing bands are also visible within this region, corresponding to rutile-TiO$_2$ and Ta$_2$O$_5$, which have been observed in Ti- and Ta-bearing superalloys exposed at similar temperatures and shown to be detrimental \cite{Wang2023TheSuperalloy}. Nb, W, and Mo are detected at lower concentrations throughout the scale, however, the APT analysis highlights the complex transition through the oxide layer. The uppermost region is dominated by CrO$^+$ ions, consistent with the chromia layer detected by STEM-EDX. Immediately beneath this, a narrow zone rich in TaO$^+$ and TaO$_2^+$ species is observed. These ions represent field-evaporation fragments of Ta$_2$O$_5$, as the molecular Ta$_2$O$_5^+$ species are too heavy to remain stable under APT analysis conditions. This confirms the presence of a thin tantalum-oxide band separating the chromia, alumina and spinel regions. 

Below this zone, a mixed Ni, Co, and refractory-rich region corresponds to the internal oxide bubble observed in STEM-EDX. At the interface between this region and the alumina, pronounced segregation of NbO$^+$ and WO$^+$ species defines a sharp interfacial boundary enriched in refractory oxides. Similar Nb- and W-oxide segregation to chromia/alumina interfaces has been reported in Ni- and Co-base alloys and attributed to the low mobility and high oxygen affinity of these elements under intermediate oxygen potentials \cite{Weng2015InfluenceSuperalloys,Ye2021InfluenceSuperalloy}. These likely form where the oxygen potential is lowest, causing Nb and W to oxidize partially and accumulate at the migrating alumina–spinel interface. Such refractory enrichment suggests potential kinetic trapping of slow-diffusing cations at the internal boundary.

Beneath this lies an Al- and O-enriched region consistent with the acicular alumina layer, within which elemental Cr$^+$ ions are also detected. The incorporation of chromium within the alumina likely arises from upward diffusion of Cr through the sub-scale oxide during its migration toward the chromia layer. Interestingly, boron is seen to segregate within the aluminum rich region. It is hypothesized here that boron could potentially play a role in stabilizing the alumina scale and increasing scale adherence.

As oxidation progresses to \SI{100}{\hour}, the XRD results indicate the development of a more stratified oxide structure. The increased intensity of spinel and Cr$_2$O$_3$ peaks suggests the formation of a duplex scale consisting of an outer (Ni,Co)-spinel region and an inner chromia-rich band. The emergence of a continuous chromia component likely marks a turning point in the oxidation process, as chromia generally exhibits lower oxygen-ion conductivity and a finer-grained morphology than spinel, both of which restrict mass transport. This transition toward chromia control coincides with the observed reduction in apparent oxidation rate and likely precedes the eventual establishment of a continuous alumina sub-layer at longer durations. Internal oxidation penetrating down grain boundaries is also observed, consistent with oxygen ingress through high-diffusivity boundary paths.

After \SI{1000}{\hour} exposure, the scale appears to reach a more stable configuration characterized by a dense, adherent oxide scale composed of Cr$_2$O$_3$, with a more continuous $\alpha$-Al$_2$O$_3$ sub-layer developed beneath it. The chromia layer remains compact and continuous. Meanwhile, the refractory-rich oxides (W, Nb, Ta) observed at earlier stages are only seen as isolated regions trapped between chromia and alumina, suggesting they are most likely residues rather than active diffusion barriers.

The WLI shows reduced surface roughness, and the depressions previously observed at grain boundaries have largely disappeared, indicating local overgrowth and the establishment of uniform scale coverage across grains. $\alpha$-Al$_2$O$_3$ reflections in XRD appear to get stronger and the weakening of the spinel peaks confirm progressive transformation of the scale towards a dual chromia–alumina scale structure. The $\theta \rightarrow \alpha$ transition of alumina involves significant volume shrinkage and defect annihilation, reducing both oxygen permeability and cation vacancy concentration \cite{Huang2021TheTransformation}. The resulting scale suppresses mass oxygen transport, most likely accounting for the observed long-term parabolic oxidation behavior.

A further notable feature is the apparent crystallographic dependence of oxidation, particularly the behaviour of the present annealing twins. WLI results revealed that regions corresponding to coherent twin bands exhibit significantly reduced oxide growth compared to the surrounding matrix, consistent with the lower surface reactivity and diffusivity expected for $\Sigma3$ boundaries. Similar effects have been reported in Ni- and Co-based superalloys and other alloy systems, where twins act as barriers to oxide nucleation or slow lateral scale growth due to their low-energy character and limited dislocation content \cite{Yamaura1999Structure-dependentAlloy,Yeh2014InvestigationsEngineering}. The reduced oxidation along these coherent interfaces may therefore contribute to the smoother long-term surface morphology observed after extended exposure, as neighboring grains overgrow and homogenize the scale topography.

The progressive surface smoothing observed by WLI ($S_a$ decreasing from \SI{0.158}{\micro\metre} to \SI{0.076}{\micro\metre}) is directly linked to the phases and chemistry of the oxide scale. The initially rough, spinel-rich surface at \SI{20}{\hour} (\textbf{Figure~2}) represents a kinetically favored transient state. Subsequently, thermodynamically stable Cr$_2$O$_3$ and $\alpha$-Al$_2$O$_3$ layers form beneath it, as confirmed by STEM-EDX. This new, adherent sub-scale undermines the original, poorly adherent spinel. Spinel spallation is therefore the most probable primary cause of the smoother surface highlighted by the reduced spinel reflections in the \SI{1000}{\hour} XRD data. The \SI{1000}{\hour} morphology thus represents this compact, stable Cr$_2$O$_3$/$\alpha$-Al$_2$O$_3$ layer rather than a relaxation of the original one. The reduction in visible grain-boundary depression further supports this, suggesting progressive overgrowth by the chromia, which blocks fast diffusion paths and lowers the overall oxidation rate down the grain boundaries. 

The oxidation process appears to involve three overlapping regimes. Initially, dominant outward cation diffusion produces a porous spinel scale and high mass gain. Subsequently, mixed diffusion (inward oxygen and lateral chromia growth) establishes a semi-protective, multilayered structure. Finally, a continuous $\alpha$-Al$_2$O$_3$ layer forms, restricting both oxygen and cation transport and yielding a parabolic rate law with improved adhesion. The alloy’s chemistry promotes early alumina nucleation without the volatility issues typical of Co-rich oxides, while Cr acts synergistically by forming a temporary chromia barrier that accelerates the densification of the scale. Refractory-rich ``bubbles'' retained at the chromia--alumina interface likely represent trapped remnants of transient oxides, marking the transition from mixed to protective scale growth.

The integrated kinetic, structural, and chemical evidence demonstrates that oxidation of this wrought Co/Ni alloy proceeds through a classical transformation sequence typical of a type II alloy: Ni/Co-spinel $\rightarrow$ external Cr$_2$O$_3$ + internal $\alpha$-Al$_2$O$_3$, culminating in a dense, continuous chromia + alumina scale. The combined use of WLI, SEM, XRD, STEM-EDX, and APT provides a complete multiscale view of this evolution—from bulk kinetics to atomic-scale segregation—revealing how a transient, porous oxide converts into a self-protecting barrier capable of sustaining long-term stability at \SI{800}{\degreeCelsius}.

\section{Conclusion}

The oxidation behavior and mechanisms of a novel wrought Co/Ni-based superalloy were characterized through complementary mass-gain kinetics, XRD, SEM/WLI, and STEM-EDX analyses, supported by local nanoscale chemical insights from APT at early stages. Isothermal oxidation tests were conducted at \SI{800}{\degreeCelsius} for \SIlist{20;100;1000}{\hour} to capture the transition from transient to steady-state oxidation. The following conclusions can be drawn:

\begin{enumerate}
\item The alloy exhibits a clear transition from rapid, near-linear oxidation at \SI{20}{\hour} to parabolic, diffusion-controlled behavior by \SI{100}{\hour} and beyond. The reduction in the apparent rate constant reflects the development of a more stable scale and a shift from outward cation to inward oxygen transport control.

\item Surface characterisation by WLI and SEM revealed a systematic smoothing of the oxide with exposure time: from ridge-like spinel growth at \SI{20}{\hour} ($S_a \approx \SI{1.6}{\micro\metre}$), to partial coalescence and shallow spallation at \SI{100}{\hour} ($S_a \approx \SI{1.1}{\micro\metre}$), and finally to a compact, adherent surface at \SI{1000}{\hour} ($S_a \approx \SI{0.7}{\micro\metre}$). The disappearance of boundary depressions indicates overgrowth and protection of grain boundaries, while WLI also shows that coherent twin bands oxidize more slowly than the surrounding matrix, consistent with the low reactivity and diffusivity associated with $\Sigma3$ interfaces. Together, these effects produce a uniform and mechanically stable surface at long exposures.

\item XRD analysis showed a sequential phase transformation from an outer (Ni,Co)(Cr,Al)$_2$O$_4$ spinel and NiO at \SI{20}{\hour}, through a duplex chromia--spinel structure at \SI{100}{\hour}, to a dense Cr$_2$O$_3$ layer with a continuous $\alpha$-Al$_2$O$_3$ sub-layer at \SI{1000}{\hour}. This progressive structural refinement accounts for the observed kinetic slowdown and long-term stability of the oxide.

\item Combined STEM-EDX and APT analyses revealed that Al and Cr selectively oxidize from the earliest stages, forming an incipient Al--O-rich transition zone that subsequently develops into a continuous alumina band beneath the chromia. APT further resolved chemical behavior within the complex scale, including CrO$^+$-rich zones corresponding to chromia, a narrow TaO$^+$/TaO$_2^+$ band indicative of Ta$_2$O$_5$, and NbO$^+$/WO$^+$ segregation defining a refractory-enriched interfacial boundary. These findings suggest that Ti- and Ta-rich oxides (rutile and Ta$_2$O$_5$) form transiently, while Nb- and W-rich oxides persist as isolated inclusions trapped between chromia and alumina after prolonged exposure. The refractory-rich ``bubbles'' thus most likely represent inert residues of the early transient regime and mark the transition from mixed to protective scale growth.

\item Overall, oxidation at \SI{800}{\degreeCelsius} follows a typical Type~II sequence of Ni/Co-spinel $\rightarrow$ Cr$_2$O$_3$ $\rightarrow$ $\alpha$-Al$_2$O$_3$. The balanced chemistry enables early alumina formation while Cr provides a transient chromia barrier that accelerates scale stability. The resulting dual chromia--alumina structure allows for improved oxidation resistance.
\end{enumerate}

%Supporting Information is available from the Wiley Online Library or from the author.

%\medskip
\textbf{Acknowledgements} \par %delete if not applicable))
The authors would like to acknowledge support from Rolls-Royce plc for their continued support and funding. MC acknowledges funding from Royal Society Tata University Research Fellowship URF/R1/201318, Royal Society Enhancement Award RF/ERE/210200EM1, and ERC CoG DISCO grant 101171966.

% References
\medskip

% Use the following code if you wish to generate your bibliography with BibTeX;
% replace the string "MSP-template" below with the name(s) of
% the BibTeX data base(s) you want to use.
% The resulting bibliography-output (the content of the .bbl file)
% must be pasted back into this file before submission.
% Please also include your BibTeX data base file(s) in your submission
% so that we can re-run BibTeX if necessary.
%
%\bibliographystyle{MSP}
%\bibliography{MSP-template}
%\newpage

\bibliographystyle{elsarticle-num} % or unsrtnat if you prefer
\bibliography{references.bib} % expects references.bib in same folder

\end{document}